\documentclass[11pt]{article}
\usepackage{geometry}                
\usepackage[T1]{fontenc}
\usepackage[english]{babel}

\usepackage[usenames,dvipsnames]{xcolor}
\usepackage{algorithm}
\usepackage[noend]{algpseudocode}
\usepackage{algpseudocode}
\usepackage{amsfonts,amsthm}
\usepackage{amsmath}
\usepackage[round]{natbib}
\usepackage{standalone}
\usepackage{import}
\usepackage{pgfplots} 
\usepackage{placeins}
\usepackage{bm}
\usepackage{setspace}
\usepackage{url}
\usepackage{graphicx}
\usepackage{tikz}
\usepackage{subcaption}
\usepackage{longtable}
\usepackage{sectsty}
\usepackage{titlecaps}

\sectionfont{\normalsize\MakeUppercase}
\title{A 1000-fold Acceleration of Hidden Markov Model Fitting using Graphical Processing Units, with application to Nonvolcanic Tremor Classification.}
\author{Marnus Stoltz$^1$ \and Gene Stoltz$^{2,3}$  \and Kazushige Obara$^4$ \and Ting Wang$^1$ \and David Bryant$^1$\\
{\small $1.$ Department of Mathematics and Statistics, University of Otago, New Zealand}\\
{\small $2.$  Council for Scientific and Industrial Research of South Africa, Pretoria, South Africa}\\
{\small $3.$  Department of Electronic Engineering, University of Johannesburg, South Africa}\\	
{\small $4.$  Earthquake Research Institute, University of Tokyo, Japan}\\
{\small $*$ Corresponding author.   {\tt david.bryant@otago.ac.nz}}
}

\providecommand{\keywords}[1]
{
	\small	
	\textbf{\textit{Keywords---}} #1
}

\begin{document}

\maketitle

\begin{abstract}
 Hidden Markov models (HMMs) are general purpose models for time-series data widely used across the sciences because of their flexibility and elegance. However fitting HMMs can often be computationally demanding and time consuming, particularly when the the number of hidden states is large or the Markov chain itself is long. Here we introduce a new Graphical Processing Unit (GPU) based algorithm designed to fit long chain HMMs, applying our approach to an HMM for nonvolcanic tremor events developed by Wang \textit{et al.}~(2018). Even on a modest GPU, our implementation resulted in a 1000-fold increase in speed over the standard single processor algorithm, allowing a full Bayesian inference of uncertainty related to model parameters. Similar improvements would be expected for HMM models given large number of observations and moderate state spaces (<80 states with current hardware). We discuss the model, general GPU architecture and algorithms and report performance of the method on a tremor dataset from the Shikoku region, Japan.
\end{abstract}
\keywords{\textit{Bayesian inference, Computational hardware, Seismology, Algorithm design.}}

\begin{center}
	\item \section{Introduction}
\end{center}

Slow slip events (SSEs), a type of slow earthquakes, play an important role in releasing strain energy in subduction zones, the region where one tectonic plate moves underneath another tectonic plate and sinks. 
It is currently understood that SSEs occur as shear slips on the bottom tip of subduction zones that transition between a fixed region above and slipping region below \citep{beroza2011slow}. Recent evidence suggest that nonvolcanic tremors are observed in close association with SSEs, however the causal relationship  between the two phenomena is not yet well understood. Classifying nonvolcanic tremors helps to better understand this link but can be time consuming when typically done by hand.

Recently, an automated procedure was developed by \citet{wang2018identifying} to classsify spatio-temporal migration patterns of nonvolcanic tremors. The procedure classifies tremor source regions into distinct segments in 2-D space using a Hidden Markov Model. The model is fitted using the Expectation Maximisation algorithm. Here we implement a Bayesian approach. However, fitting the model in either a frequentest or Bayesian framework is extremely demanding computationally, often taking days or weeks for large dataset with moderate state space. Fortunately, technological advances in hardware have the potential to solve this issue. Specifically, we make use of fast and affordable graphic processing units (GPUs).

In recent years HMM algorithms on GPUs have been implemented in various fields. A non-exhaustive list includes implementations in bioinformatics~\citep{Yao2010}, speech recognition~\citep{Yu2015}, a registered patent in speech matching~\citep{chong2014utilizing} and workload classification~\citep{Cuzzocrea2016}, as well as HMMer~\citep{Horn2005} an open-source project for use with protein databases. The HMM implementations are application specific often with large number of states and mostly focused on increasing throughput of the Verterbi and Baum-Welch algorithms \citep{Zhang2009, Li2009, Liu2009}. This leads to a range of concurrent approaches. Here we focus on the efficient implementation of the forward algorithm of an HMM model given a large number of observations and a moderate number of states. 

The outline of the paper is as follows: In Section 2 we describe the HMM for classifying nonvolcanic tremors and discuss the likelihood algorithm in a serial and parallel context. Thereafter we give details on the OpenCL implementation of the parallel likelihood algorithm. In Section 3 we discuss performance of the OpenCL implementation and compare it to the standard Forward algorithm. In Section 4 we report our analysis on a large tremor dataset from the Shikoku region, Japan. 

\begin{center}
\item \section{An HMM for classifying nonvolcanic tremors}
\end{center}
Nonvolcanic tremor activity is clustered spatially and each spatial cluster seems to recur episodically. To represent this phenomenon using an HMM, \citet{wang2018identifying} introduce one hidden state for each spatial cluster. The tremors themselves (including the absence of a tremor) are the observations. The frequency and spatial distribution of tremors changes according to the hidden state. 

More formally, we suppose that the observations of nonvolcanic tremors are a sample path of a stochastic process \[\{X_i\}_{i=0,\dots,N}\] 
with observations represented in  the state space
\[I = \{\emptyset, \mathbb{R}^2\} \]
generated under an 
HMM with $K$ numbered hidden states. For each hidden state $k=1,\dots,K$ we introduce parameters $p_{k}$, $\bm{\mu}^{(k)}$ and $\mathbf{\Sigma}^{(k)}$, where $p_{k}$ is the  probability of observing a tremor  and $\bm{\mu}^{(k)}$, $\mathbf{\Sigma}^{(k)}$  are the mean and variance of a bivariate normal distribution modelling where a tremor is likely to occur, if it does occur.    

To simplify notation we introduce for each observation $\mathbf{x}$ a $K \times K$ diagonal matrix $\mathbf{P}(\mathbf{x})$, also called the \textit{emission matrix}, with the $k$th diagonal element corresponding to the probability of observing $\mathbf{x}$ given state $k$
\begin{equation} 
\bm{P}(\bm{x})_{kk} = 
\begin{cases} 
p_{k} \phi(\bm{x}|\bm{\mu}^{(k)},\mathbf{\Sigma}^{(k)}) \\
1-p_{k}. 
\end{cases}
\label{eqn: emission-matrix}
\end{equation}
Here $\phi(.)$ is the density function of bivariate normal distribution. 
Let $\mathbf{\Gamma}=(\Gamma_{ij})$ denote the $K \times K$ transition matrix of the HMM, where $\Gamma_{ij}$ indicate the the transition probability from hidden state $K=i$ to $K=j$. Also, let $\bm{\delta} = \delta_1,\dots, \delta_K$ denote the vector of probabilities for the initial state. 

Now the likelihood function for the parameters given the observed data can be written as  
\begin{equation}
\mathrm{L}\big(\mathbf{\Gamma}, \bm{\delta}, \{p_k,\bm{\mu}^{(k)},\mathbf{\Sigma}^{(k)}\}_{k=1,\ldots,K} | \mathbf{x}_0,\ldots,\mathbf{x}_N \big) =  \bm{\delta}^T \mathbf{\Gamma}\mathbf{P}(\mathbf{x}_0)  \dots \mathbf{\Gamma}\mathbf{P}(\mathbf{x}_N)  \mathbf{1}.  
\label{eqn: likelihood func}
\end{equation}

\begin{center}
	\item \section{GPU computing framework}
\end{center}

 GPUs have had a large impact across statistical and computing sciences due to cost-effect parallelism \citep{kindratenko2014numerical}.
However in order to translate an algorithm from CPU to GPU some careful consideration is needed in terms of
\begin{enumerate}
	\item Reducing latency  (how to concurrently execute instructions on GPU in order to optimise data throughput.)
	\item Managing memory (how to effectively distribute and utilise memory across processors to avoid bandwidth bottlenecks).
	\item Designing robust algorithms with respect to varying GPU architecture between models and vendors as well as the rapidly changing landscape of computational hardware.
\end{enumerate}
Frameworks like OpenCL and CUDA, allow programmers to implement GPU algorithms with some level of generality. 
The implementation we describe here was carried out in the OpenCL framework. OpenCL is an open standard maintained by the non-profit technology consortium Khronos Group, see \url{https://www.khronos.org} for more details on the non-profit organisation. 

The OpenCL framework consists of a \textit{host} (CPU; terms in brackets relate to computation on GPU architecture) controlling one or more \textit{compute devices} (We just used one GPU). Each compute device (GPU) is divided into \textit{compute units} (streaming multiprocessors). Compute units are further divided into \textit{compute elements} (microprocessors or cores). Each compute unit has access to global memory of the compute device. This access though is slow. Each compute unit also has a shared memory to allow efficient data exchange between compute elements. Each compute element has exclusive access to private memory (registers) for computation. 

\begin{center}
	\item \section{The likelihood algorithm}
\end{center}
\subsection{Overview}
Our implementation will work well on a range of GPU models. For our studies we used a NVIDIA GeForce GTX 1080 Ti GPU with 28 compute units (streaming multiprocessors) each with 48KB of shared memory, 128 compute elements (cores)  and a register file that can contain up to 32,768 32-bit elements distributed across the compute elements (cores). For the host we used an Intel Core i7-7700K CPU at 4.20GHz. 
There are two main limitation for the OpenCL algorithm in terms of hardware specifications 
\begin{enumerate}
	\item 	The number of registers per compute element.
	\item   The size of shared memory on a compute unit. 
\end{enumerate} 
For example given the hardware described above we have (32,768/128)=256 registers per compute element. This implies that we can store up to roughly 200 32-bit matrix elements on a compute element (we also need some registers left to store counters and other meta variables). Our implementation assumes that at least two matrix rows can fit into the registers of a compute element. This gives an upper limit for the number of hidden states of $K<100$. In order to efficiently distribute rows of a matrix and update matrix elements we need space for two matrices in the shared memory of the compute unit. Our configuration has 48KB of shared memory per compute unit. Implying that we can fit a total of $(48\cdot2^{10})/4=12288$ 32-bit matrix elements per compute unit. This gives a second upperlimit for number of hidden states of $K<80$. To handle a large number of states, alternative parallel computing strategies should be used \citep{Horn2005, Yu2015}. 

First we consider how the algorithm for the likelihood \eqref{eqn: likelihood func} would be implemented on a single processor unit. To avoid matrix-matrix multiplications we would start with the stationary vector $\bm{\delta}$ on the left, and then sequentially multiply that by transition matrices and emission matrices:

\begin{algorithm}
	\caption{The Forward algorithm on a CPU}
	\label{alg:forward}
	\begin{algorithmic}[1]
		\Procedure{Compute-likelihood}{$\{p_k,\bm{\mu}^{(k)},\mathbf{\Sigma}^{(k)}\}_{k=1,\ldots,K}, \{\mathbf{x}_0,\dots,\mathbf{x}_N\}$}
		\State $\bm{v} \gets \bm{\delta}^T$
		\For{$k$ from $0$ to $N$}
		\State Compute $\bm{P}(x_k)$
		\State $\bm{v} \gets \bm{v}\bm{\Gamma} $
		\State $\bm{v} \gets \bm{v}\bm{P}(x_k) $
		
		\EndFor
		\State \Return $\bm{v}\bm{1}$
		\EndProcedure
	\end{algorithmic}
\end{algorithm}

Running time will be dominated by the matrix-vector multiplication in steps 5 and 6, taking $\mathcal{O}(K^2)$ time per iteration. Hence the running time, or work, for this implementation is $\mathcal{O}(NK^2)$. Next we compare it with the parallel implementation. 

The overview of our implementation is as follows: 
\begin{enumerate}
\item We compute all of the emission matrices $\bm{P}(x_0),\ldots \bm{P}(x_N)$ in parallel 
\item We then multiply the emission matrices by the transition matrices, all in parallel storing N matrices   $ \bm{\Gamma} \bm{P}(x_0) , \dots ,\bm{\Gamma}\bm{P}(X_N)$. 
\item Instead of computing 
$\bm{\Gamma}\bm{P}(x_0) , \dots ,\bm{\Gamma}\bm{P}(X_N)$
as a single sequence of vector-matrix multiplications, we multiply the matrices
$( \bm{\Gamma}\bm{P}(x_0)), \dots ,(\bm{\Gamma}\bm{P}(X_N))$
together.	
\end{enumerate}
 This {\em increases} the work done: we are carrying out matrix-matrix multiplications instead of matrix-vector multiplications, but it allows us to spread the computation over multiple processors. We now discuss steps (1) to (3) in greater detail.

\subsection{Step 1: The emission probability evaluation on GPU}
The goal in this step is to compute the emission matrices $\mathbf{P}(x_i)$ for each observation $x_i$. The emission probability is defined by \eqref{eqn: emission-matrix} and makes use of the parameters  $p_k, \bm{\Sigma}_k, \bm{\mu}_k$ for each hidden state $k$. These parameters are initially copied to the registers of each core and remains there until all the datapoints have been evaluated.  
The compute elements work in parallel. Each is allocated a data point $x_i$, uses the stored values to compute $\bm{P}(x_i)$ and copies the diagonal matrix computed to global memory. Note that a compute element can request and copy the next data point at the same time as it processes the current data point. 

For this step there is no data sharing between compute elements, allowing for data-level parallelism. Therefore it is more efficient to allow compute device compiler to optimise the work-load scheduling and data transfer between compute units in order to fully utilise SIMD (Single instruction multiple data) instructions. Output from compute elements are collected and copied to global memory to form the list of new inputs $\{\mathbf{\Gamma}, \mathbf{P}(x_0),\dots,\mathbf{P}(x_N)\}$ for the next kernel.

\subsection{Step 2: The transmission-emission matrix multiplication on GPU}
During the next step we compute $\bm{\Gamma} \bm{P}(x_i)$ for all data points $x_i$, again in parallel. At this point we run into limitations with memory. While the register of a single compute element is large enough to store the diagonal matrix $\bm{P}(x_i)$, it is not large enough to store the full transition matrix $\bm{\Gamma}$ nor the product matrix $\bm{\Gamma} \bm{P}(x_i)$. The solution is to break down the multiplication of $\bm{\Gamma}$ and $\bm{P}(x_i)$ by computing only a few rows at once. 

We query the register size for each compute element to determine how many rows of $\mathbf{\Gamma}$ can be copied. The rows remain in the register until all data points have been evaluated. Thereafter the next set of rows is copied into the registers and the data points is evaluated again until all the rows of $\mathbf{\Gamma}\mathbf{P_r}$ for $r=0,\dots,N$ have been computed.    
As $\bm{P}(x_i)$ is diagonal, the product of rows of $\bm{\Gamma}$ with $\bm{P}(x_i)$ is computed by simply rescaling the corresponding columns.

The next diagonal matrix subset is requested while scaling subset for current data point.
Again, there is no data sharing between compute elements, allowing for optimal data-level parallelism. Output from compute elements are collected and a new list of inputs, namely $\{(\mathbf{\Gamma P_0}),\dots,(\mathbf{\Gamma P_N})
\}$ is compiled for the final GPU kernel.

%

\subsection*{Step 3: The Square Matrix-Chain Multiplication on GPU}
The third step is the most time-consuming, and also the most involved. The general idea is to avoid the long sequence of matrix vector calculations
\[ \bm{\delta}^T \mathbf{\Gamma}\mathbf{P}(x_0)  \dots \mathbf{\Gamma}\mathbf{P}(x_N) \bm{1} \]
which cannot be readily parallelized, 
by instead multiplying the matrices together in parallel. Our algorithm here roughly follows \citep{Masliah:2016:Matrices}. 

Recall the general hierarchical structure of a GPU calculations, as described above. 
The CPU controls the GPU. Each compute GPU is divided into compute units (streamline multiprocessors). Compute units are further divided into compute elements (microprocessors or cores). The CPU is actually faster than the compute units for individual computations, the speed of GPUs being due to parallelism. Our algorithm takes advantage of all three levels: The sequence of matrices (known in computer science as a {\em matrix chain}) is divided into multiple segments, one for each compute unit. The compute units then carry out matrix multiplication directly, making use of multiple compute elements to share out the rows in each matrix-matrix computation. We then use the CPU to carry out the final sequence of matrix-vector computations, using the matrices returned by the computational units of the GPU. 

Note that in practise we compute $\log$L rather than L and shift the registers either up or down using the scale coefficients from compute units to avoid underflow.

\FloatBarrier
\begin{center}
	\item \section{Performance assessment of OpenCL implementation}
\end{center}

\subsection{OpenCL algorithm vs Forward algorithm}
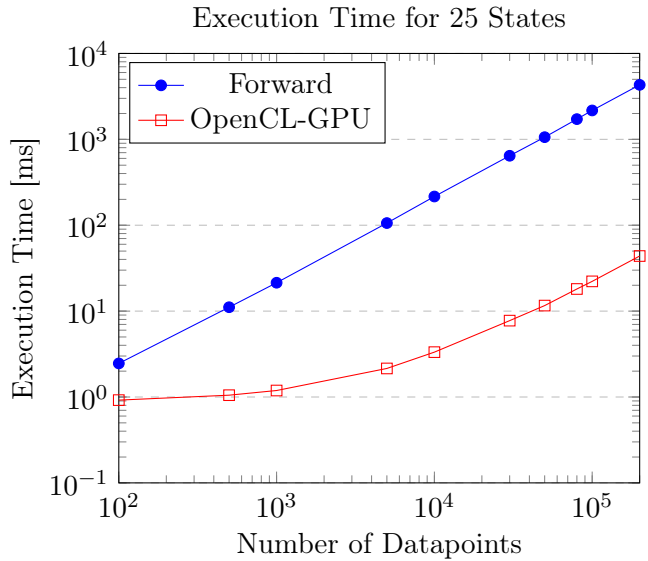
\begin{figure}
	\centering
	    \begin{tikzpicture}
        \begin{loglogaxis}[
            cycle list={
            {blue,mark=*},
            {red,mark=square},
            {dashed,mark=o},
            {loosely dotted,mark=+},
            {brown!60!black,mark=otimes*,mark options={fill=brown!40}}
            },
            title={Execution Time for 25 States},
            xlabel={Number of Datapoints},
            ylabel={Execution Time [ms]},
            legend style={at={(0.02,0.98)},anchor=north west},
            xmin=100, xmax=200000,
            ymin=0.1, ymax=10^4,
            ymajorgrids=true,
            grid style=dashed,
        ]
        
            \addplot table {

         100     2.46
         500     11.1
        1000     21.4
        5000    1.06e+02
       10000    2.16e+02
       30000    6.44e+02
       50000    1.06e+03
       80000    1.72e+03
      100000    2.17e+03
      200000    4.31e+03

            };           

            \addplot table {

         100     0.92
         500     1.05
        1000     1.19
        5000     2.15
       10000     3.33
       30000     7.75
       50000     11.6
       80000     18.1
      100000     22.2
      200000     43.8

            };

            \legend{Forward, OpenCL-GPU}
        \end{loglogaxis}
    \end{tikzpicture}
	\caption{We compare computational time of OpenCL algorithm on GPU with a Forward algorithm on CPU. Computational time is indicated on the y-axis and number of datapoints are indicated by the x-axis. We see that with $10^5$ datapoints, the GPU algorithm runs $\sim 10^3$ times faster. }
	\label{fig:graph_clperf}
\end{figure}

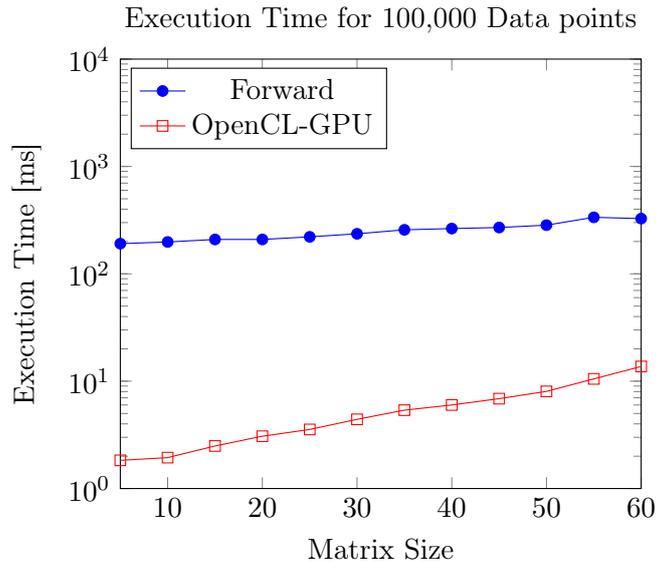
\begin{figure}
	\centering
	    \begin{tikzpicture}
        \begin{semilogyaxis}[
            cycle list={
            {blue,mark=*},
            {red,mark=square},
            {dashed,mark=o},
            {loosely dotted,mark=+},
            {brown!60!black,mark=otimes*,mark options={fill=brown!40}}
            },
            title={Execution Time for 100,000 Data points},
            xlabel={Matrix Size},
            ylabel={Execution Time [ms]},
            legend style={at={(0.02,0.98)},anchor=north west},
            xmin=5, xmax=60,
            ymin=1, ymax=10^4,
        ]
              \addplot table {
           5    1.91e+02
          10    1.98e+02
          15    2.09e+02
          20    2.09e+02
          25    2.21e+02
          30    2.36e+02
          35    2.57e+02
          40    2.64e+02
          45    2.7e+02
          50    2.84e+02
          55    3.36e+02
          60    3.27e+02
          65    3.92e+02
          70    3.94e+02
          75    4.1e+02
                };


            \addplot table {
           5     1.83
          10     1.94
          15     2.49
          20     3.07
          25     3.55
          30     4.41
          35     5.37
          40      6.0
          45     6.87
          50     8.03
          55     10.5
          60     13.7
          65     15.1
          70     20.4
          75     19.4
            };            
              
            \legend{Forward, OpenCL-GPU} 
        \end{semilogyaxis}
    \end{tikzpicture}
	\caption{We compare computational time of OpenCL algorithm on GPU with a Forward algorithm on CPU. Computational time is indicated on the y-axis and number of HMM states are indicated by the x-axis. We see that the GPU algorithm slows down as the register capacity of compute elements is reached. However it still outperforms the Forward algorithm by orders of magnitude. }
	\label{fig:graph_clperf_2}
\end{figure}

One of the factors that influence the use of an algorithm on GPUs is whether it is actually faster than a Forward algorithm. To check this we compare computational times of the GPU algorithm with the Forward algorithm from the software library Tensor flow. First we fixed the number of HMM states to $K=25$ while increasing the number of datapoints over a range of magnitude orders $N=10^2,\dots,10^5$. Thereafter we fixed the number of datapoints to $N=100,000$ and increased the number of HMM states for $K=5,10,\dots,50$. In each case model parameters were drawn from the prior disribution (discussed in the next section) and thereafter data was simulated using the R software package in \citet{wang2018identifying}. The results are shown in Figure~\ref{fig:graph_clperf} and Figure~\ref{fig:graph_clperf_2}. We see that the GPU algorithm executes orders of magnitude faster then a Forward algorithm.

\subsection{Comparing execution time of matrix-chain multiplication}
\begin{figure}
	\centering
	    \begin{tikzpicture}
        \begin{loglogaxis}[
            cycle list={
            {blue,mark=*},
            {red,mark=square},
            {dashed,mark=o},
            {loosely dotted,mark=+},
            {brown!60!black,mark=otimes*,mark options={fill=brown!40}}
            },
            title={Speed Comparison for 50 States},
            xlabel={Data Points},
            ylabel={Execution Time [ms]},
            legend style={at={(0.02,0.98)},anchor=north west},
            xmin=100, xmax=200000,
            ymin=0.1, ymax=10000,
            ymajorgrids=true,
            grid style=dashed,
        ]
            \addplot table {
         100    21.833499
         500    40.064946
        1000    57.829031
        5000    192.144602
       10000    336.950871
       30000    1022.257873
       50000    1576.167799
       80000    2544.715574
      100000    3161.085144
      200000    6200
            };
            \addplot table {
         100    0.155926
         500    0.477076
        1000    0.997066
        5000    3.306150
       10000    7.332087
       30000    18.092871
       50000    34.177780
       80000    59.775829
      100000    93.347788
      200000    155.735731
            };            
            \addplot table {
         100    1.119614
         500    1.322937
        1000    1.955628
        5000    4.395866
       10000    6.682873
       30000    15.663266
       50000    24.476337
       80000    39.353275
      100000    50.345898
      200000    95.492744

            };    
            
            \legend{CLBlast, MAGMA, OWN}
        \end{loglogaxis}
    \end{tikzpicture}%
	\caption{For this computational comparison (in miliseconds) with the BLAS libraries we fix the number of HMM states to $K=50$ and increase the number of datapoints over a range of magnitude orders.}
	\label{fig:graph_speedseq}
\end{figure}
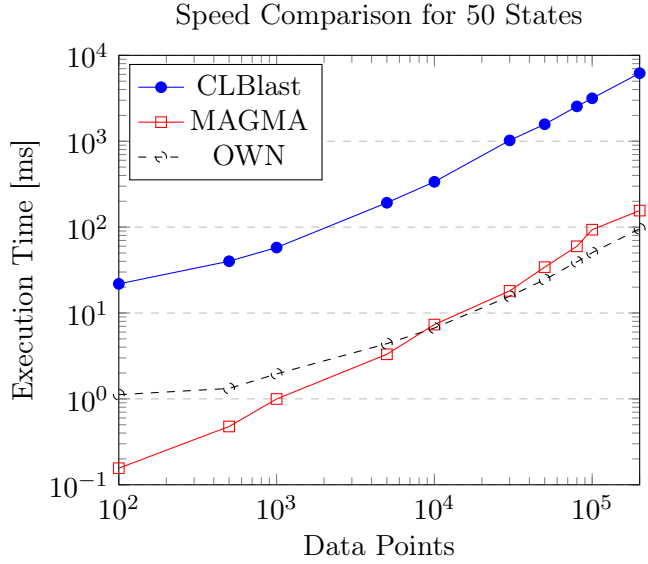

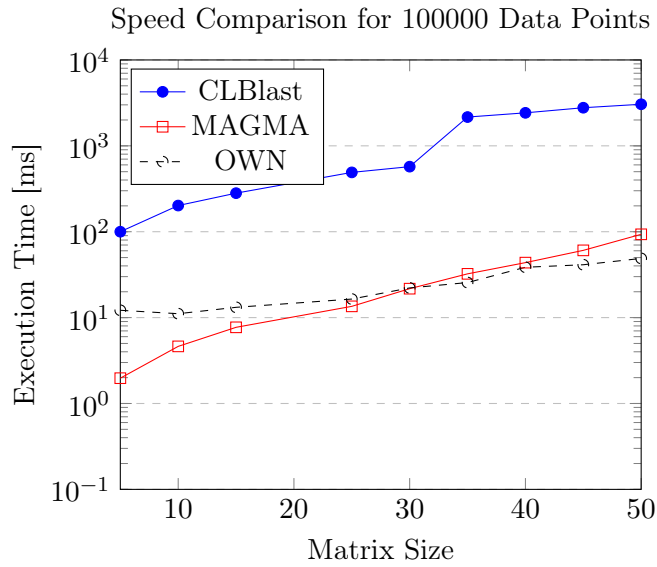
\begin{figure}
	\centering
	    \begin{tikzpicture}
        \begin{semilogyaxis}[
            cycle list={
            {blue,mark=*},
            {red,mark=square},
            {dashed,mark=o},
            {loosely dotted,mark=+},
            {brown!60!black,mark=otimes*,mark options={fill=brown!40}}
            },
            title={Speed Comparison for 100000 Data Points},
            xlabel={Matrix Size},
            ylabel={Execution Time [ms]},
            legend style={at={(0.02,0.98)},anchor=north west},
            xmin=5, xmax=50,
            ymin=0.1, ymax=10000,
            ymajorgrids=true,
            grid style=dashed,
        ]
            \addplot table {
      5    100
     10    201.677606
     15    281.159546
     25    490.631568
     30    572.241467
     35    2163.224209
     40    2422.104395
     45    2777.759543
     50    3048.026727
            };
            \addplot table {
      5    1.971006
     10    4.622936
     15    7.707834
     25    13.523817
     30    21.713972
     35    32.322168
     40    43.655157
     45    60.670137
     50    93.284130
            };            
            \addplot table {
      5    12.168217
     10    11.101389
     15    13.173699
     25    16.441607
     30    22.157526
     35    25.591326
     40    38.564944
     45    41.251278
     50    48.858595
            };    
            
            \legend{CLBlast, MAGMA, OWN}
        \end{semilogyaxis}
    \end{tikzpicture}
	\caption{For this computational comparison (in miliseconds) with the BLAS libraries we fix the number of datapoints to $N=100,000$ and increase the number of HMM states for $K=5,10,\dots,50$.}
	\label{fig:graph_speedmatrix}
\end{figure}

Here we specifically compare computation time of step 3 in the OpenCL algorithm with matrix-chain multiplication using popular GPU BLAS (Basic Linear Algebra Subprograms) libraries. We use subroutines from the CLBlast library as well as the MAGMA BLAS libary to do the matrix-chain multiplication. CLBLast is a general BLAS library in OpenCL that automatically tunes subroutines for specific hardware based on compile time. MAGMA BLAS is a CUDA library exclusively available for NVIDIA GPUs. We followed the same procedure as in the previous two experiments except that we fixed the number of HMM states to $K=50$. We show results in Figure~\ref{fig:graph_speedseq} and Figure~\ref{fig:graph_speedmatrix}. Using the MAGMA library gives roughly the same performance as the OpenCL algorithm for small matrices. We note that using these libraries in the OpenCL algorithm is not straightforward due to small tweaks and scaling coefficients that we keep track of in addition to performing the matrix-chain multiplication.  
The algorithm became very slow when the HMM had more than 100 states due to memory limitations previously discussed. 

\begin{center}
	\section{Bayesian analysis of nonvolcanic tremor data}
\end{center}

\subsection{Monte Carlo Markov Chains}
Bayesian techniques have become a popular method of statistical inference across a broad range of sciences \citep{johannesson2016bayesian, kruschke2010believe, moore2005internet, stoltz2019bayesian, turner2016bayesian, woolrich2009bayesian}. This is due to advances in numerical techniques and the affordability of powerful computers \citep{andrieu2004computational}. In a Bayesian analysis the aim is to compute the joint posterior distribution of model parameters, simply referred to as the posterior distribution. The posterior distribution summarizes the uncertainty related to model parameters. 

Typically due to model complexity the posterior distribution is an analytically intractable function. However methods such as Monte Carlo Markov Chains (MCMCs) use random walks to estimate the posterior distribution. The basic concept behind MCMCs is that a Markov chain can be constructed with a stationary distribution, where the stationary distribution is in fact the posterior distribution. 

An MCMC is initialized by choosing a random state, typically by drawing a sample from the prior distribution (we discuss prior distributions below).
The MCMC is then simulated by accepting or rejecting proposed MCMC states based on a ratio of the likelihood function and prior distribution of both the current and proposed MCMC state. The MCMC is simulated until after the stationary distribution is reached. Stationarity of an MCMC is assessed by looking at trace plots of parameters as well as computing the number of effective independent samples. Samples of the stationary MCMC is then used to approximate the posterior distribution. 

\subsection{Model priors}
Using ratios of the likelihood and prior distributions is an elegant way of sampling from the
posterior distribution. It sidesteps some nasty calculations if we were to compute the posterior 
distribution directly instead. Roughly speaking prior distributions is a way to incorporate knowledge about model parameters before looking at the data. However prior distributions can easily be neglected but they are in fact an important part of the model. Therefore choosing prior distributions needs to be carefully considered and requires some justification.  For instance, it is known that tremors occur in sequence bursts that cluster around the same area \citep{wang2018identifying}. This observation we translate into the model by specifying a model prior centred around sparse transition matrices. More formally, we specify a symmetric Dirichlet prior with concentration parameter 0.01 on  $\bm{\Gamma}$ (formulas for prior densities are given in Appendix A). Furthermore we expect that for some hidden states we are more likely to observe tremors than others. Therefore we specify independent gamma distributions on state probabilities $\{p_k\}_{k=1,\ldots,K}$ , half of the state probabilities with mean 0.1 and variance 0.001 and the other half with mean 0.9 and variance 0.001. Also, we specify a uniform prior on hidden state means $\{\bm{\mu}^{(k)}\}_{k=1,\ldots,K}$ restricted to a rectangular domain that contains all observations. We have no prior information on the shape of the hidden states therefore we specify an uninformative Inverse-Wishart prior on the covariance matrices $\{\mathbf{\Sigma}^{(k)}\}_{k=1,\ldots,K}$ with degrees of freedom equal to the number of states $K$ and scale matrix set to a $K \times K$ identity matrix.      

\subsection{GPUeR-hmmer}
In order to simulate MCMCs for the model, we incorporated the GPU likelihood algorithm along with the prior distributions into a 
general purpose MCMC sampler \citep{christen2010general}. This R package bundle is freely available at \url{https://github.com/genetica/HMMTremorRecurrencePatterns}. Note that OpenCL 1.2 and Python 3.6 (or later versions) needs to be separately installed on a system in order to support the back-end of the R package. The R package also contains a simple example using simulated data from the HMM described in Section 2. Additionally we provide instructions on how to modify the OpenCL code if an HMM with a different emission function is required. In order to assess convergence of the MCMC chains we used Tracer. For a brief tutorial on how to use Tracer to assess convergence, see \url{https://beast.community/analysing_beast_output}. Furthermore if some problems are encountered with convergence of the simulated MCMCs see \url{https://beast.community/tracer_convergence} for some recommendations. 

\subsection{Tremor dataset of the Shikoku region}
 We use a large tremor dataset from the Shikoku region, Japan to demonstrate the sort of Bayesian analysis that can be done with GPUeR-hmmer. The Shikoku region is one the three major regions in Japan (the other two being the Tokai region and Kii region) in which nonvolcanic tremor occurrences have been repeatedly detected. Tremor activity spans along the strike of the Phillipines Sea plate for about 600km and the depth ranges from 30 to 45 km on the plate interface. The original waveform data is supplied by the High Sensivity Seismograph Network of the National institute for Earth Sciences and Disaster prevention in Japan. The dataset analysed by \citet{wang2018identifying} was extracted from the waveform data. It consists of $105,000$ data point measurements between $2001$ and $2012$. It is hourly control measurements determined using clustering and correlation methods described in \citet{obara2010depth}. 

\subsection{Model fitting}
A full Bayesian analysis of the model will sample the number of hidden states along with the rest of the model parameters. However sampling from different parameter spaces is quite challenging and is an active and ongoing area of research  \citep{lunn2009generic}. Instead we incorporate the choice of number of hidden states $K$ into the model fitting process.

We start with a small number of hidden states and incrementally increase the number of hidden states, while doing so we assess the posterior distribution for each case. The posterior distribution of each model is estimated by running the MCMC sampler for 1,000,000 iterations.  Running each chain took approximately $\sim 4$ hours. 

In Figure \ref{fig:map} we summarize the posterior distributions for model fitted with number of hidden states $K=5,10, \dots, 30$. Typically, the background states (i.e states that cover large areas) have the highest variance in posterior distribution. Whereas states covering smaller areas have considerably less variance in posterior distribution of parameters. We also see in Figure~\ref{fig:map}(d) that parameters used in \citet{wang2018identifying} are recovered by the posterior distribution. Typically as we increase the number of states some states are divided into two, with rare new clusters. Furthermore we see for $K=30$ that some additional hidden states ($k=4,8,26$) doesn't fit over one particular cluster of points, covers a large area, has a low probability of observing tremors and a low stationary probability (i.e time spent in state). Thereafter we also fitted models with hidden states for $K=26,27$ (see Appendix B) and we find that additional hidden states have the same undesirable properties and therefore use $K=25$ as our choice for number of hidden states for the model (see MCMC summary statistics in Appendix C).  
     
\begin{figure}
	\centering
	\begin{subfigure}{\columnwidth}
		\begin{tikzpicture}
		\node(a){\includegraphics[width=\columnwidth]{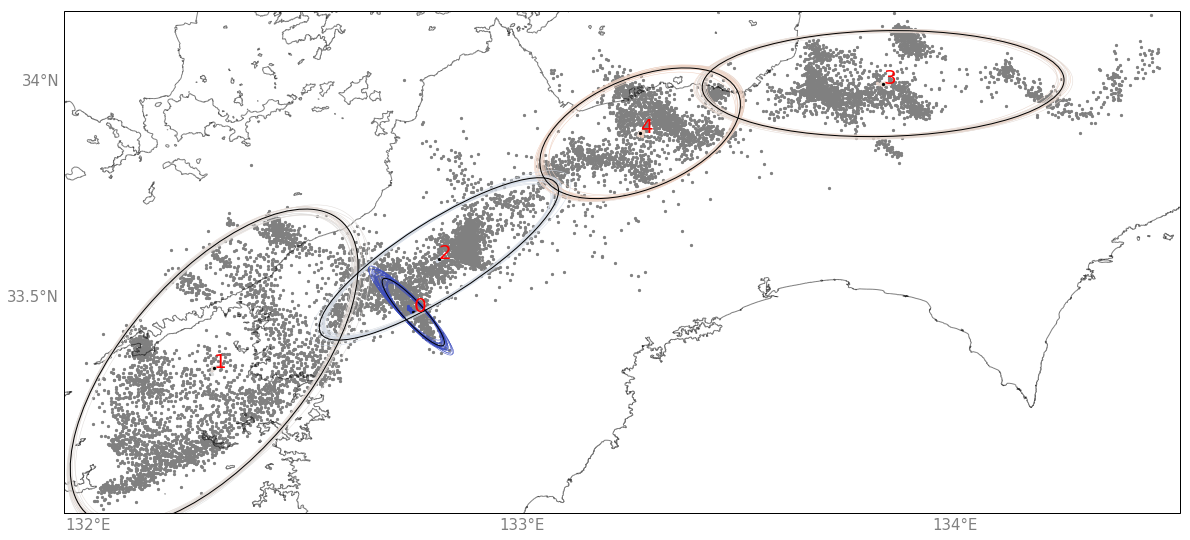}};
		\node at (a.south east)
		[
		anchor=center,
		xshift=-20mm,
		yshift=20mm
		]
		{
			\includegraphics[width=0.25\columnwidth]{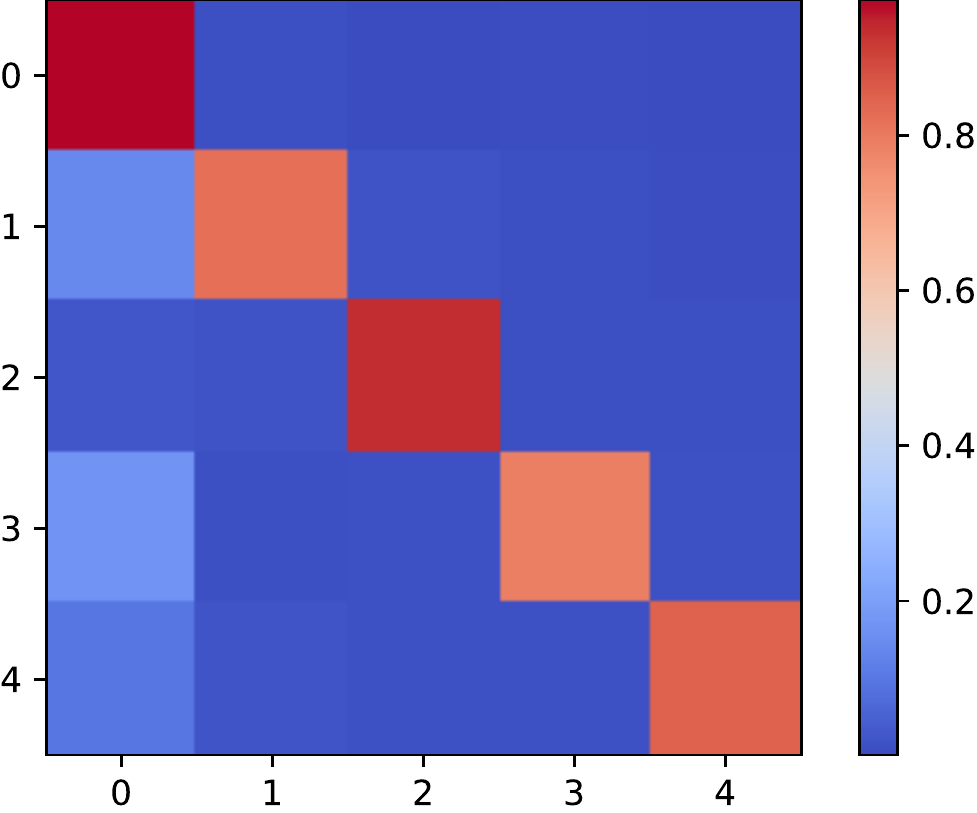}
		};
		
		\end{tikzpicture}%
		\caption{$K=5$}
	\end{subfigure}
	
	\begin{subfigure}{\columnwidth}
		\begin{tikzpicture}
		\node(a){\includegraphics[width=\columnwidth]{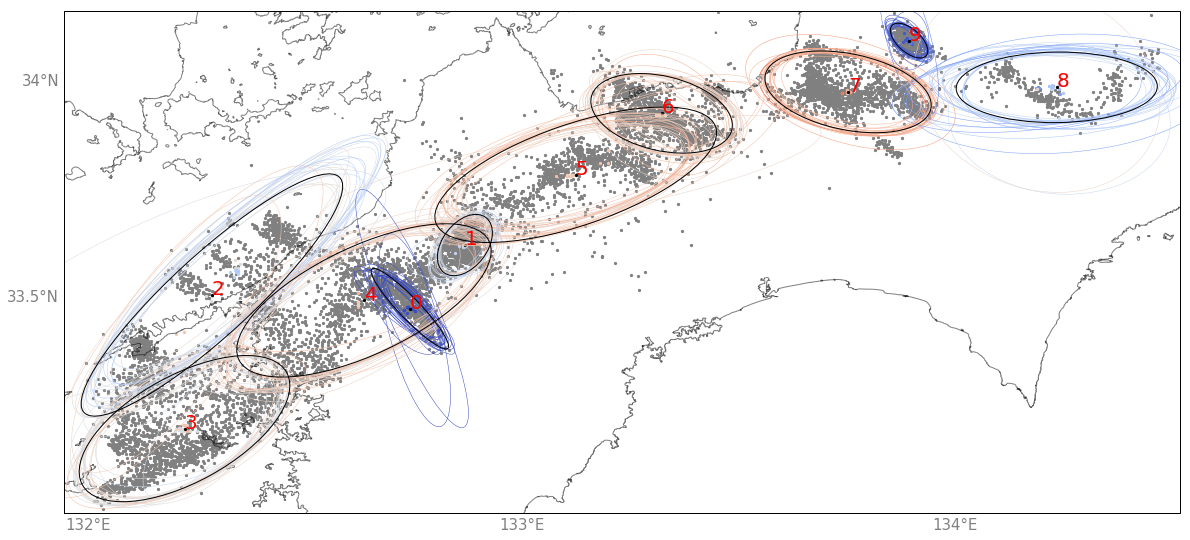}};
		\node at (a.south east)
		[
		anchor=center,
		xshift=-20mm,
		yshift=20mm
		]
		{
			\includegraphics[width=0.25\columnwidth]{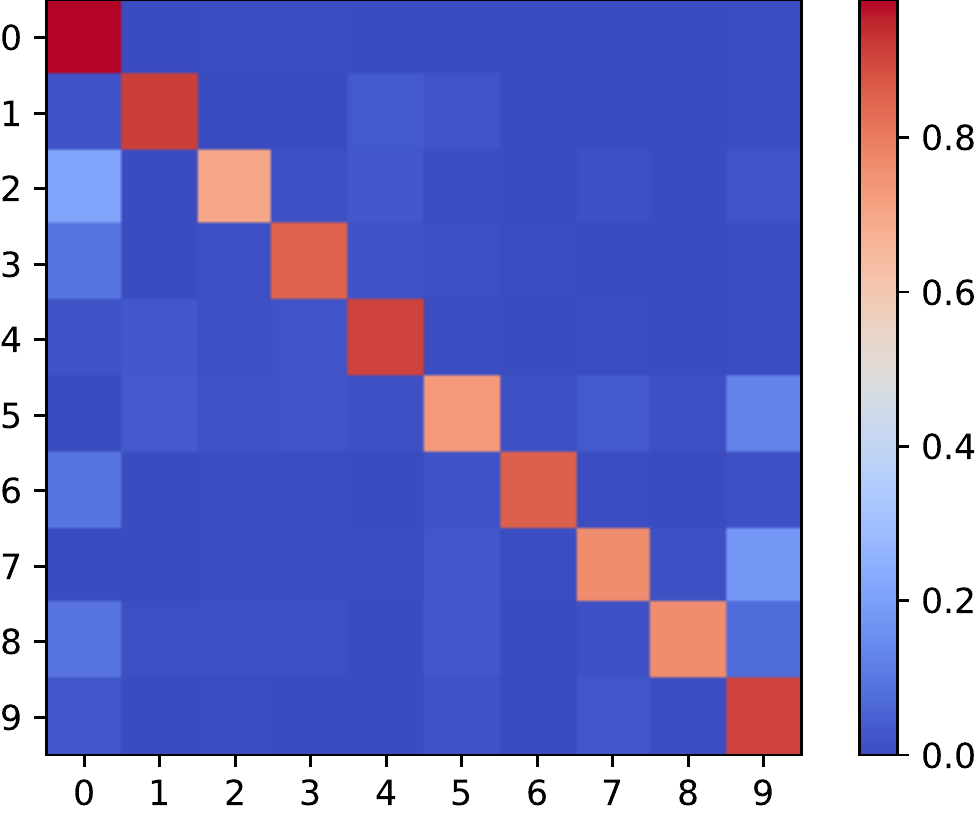}
		};
		
		\end{tikzpicture}%
		\caption{$K=10$}
	\end{subfigure}
\end{figure}

\begin{figure}
	\ContinuedFloat
	\begin{subfigure}{\columnwidth}
		\begin{tikzpicture}
		\node(a){\includegraphics[width=\columnwidth]{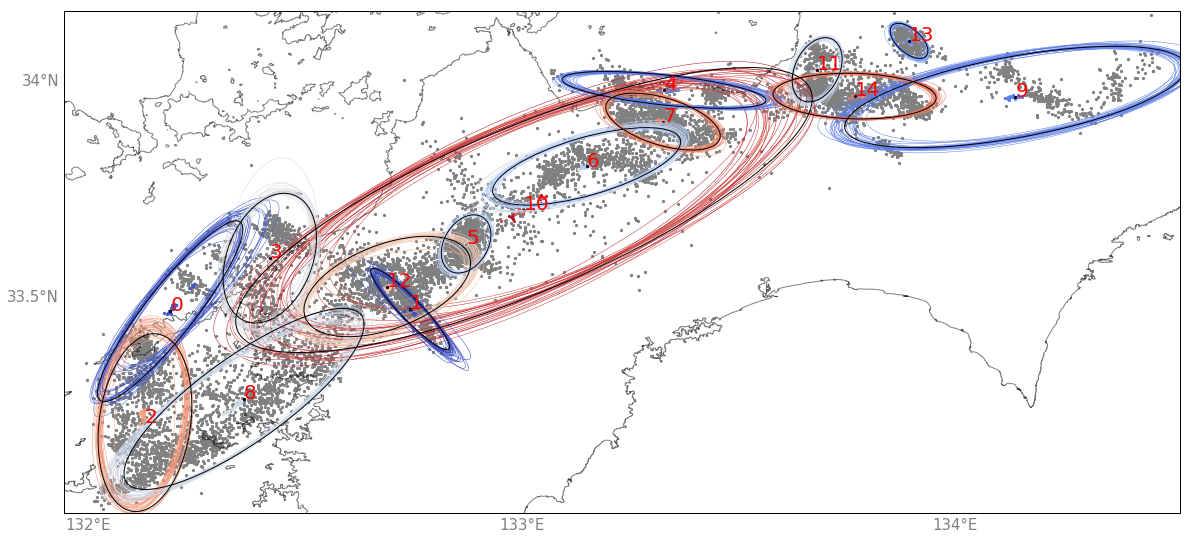}};
		\node at (a.south east)
		[
		anchor=center,
		xshift=-20mm,
		yshift=20mm
		]
		{
			\includegraphics[width=0.25\columnwidth]{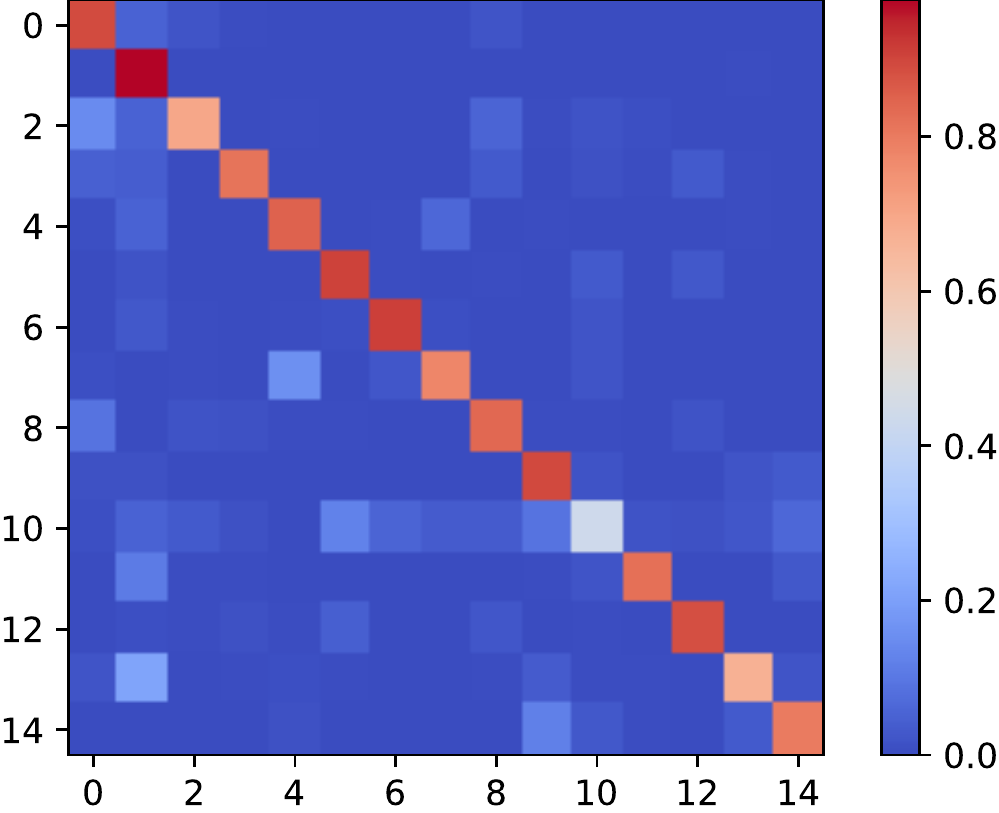}
		};
		
		\end{tikzpicture}%
		\caption{$K=15$}
	\end{subfigure}
	
	\begin{subfigure}{\columnwidth}
		\begin{tikzpicture}
		\node(a){\includegraphics[width=\columnwidth]{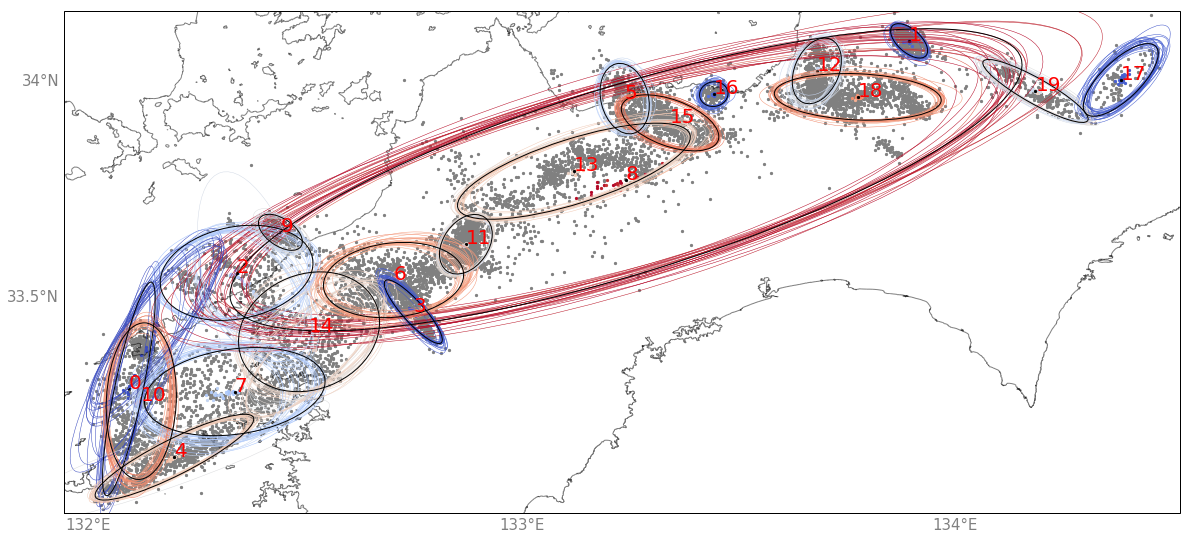}};
		\node at (a.south east)
		[
		anchor=center,
		xshift=-20mm,
		yshift=20mm
		]
		{
			\includegraphics[width=0.25\columnwidth]{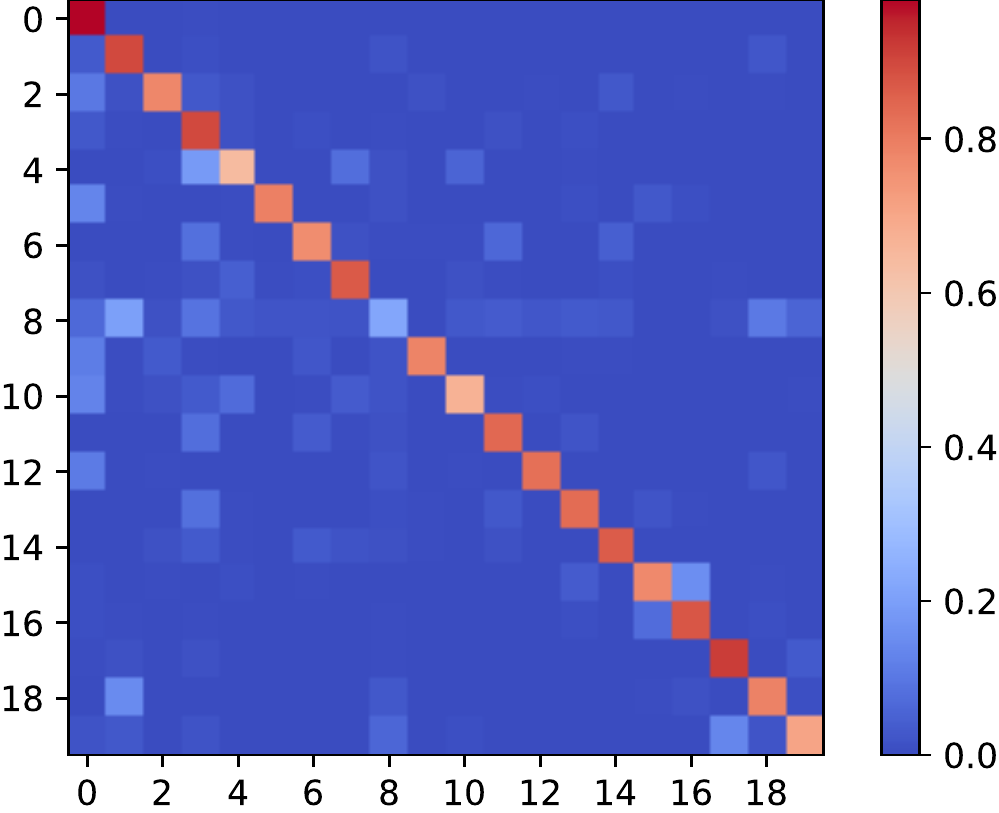}
		};
		
		\end{tikzpicture}%
		\caption{$K=20$}
	\end{subfigure}
\end{figure}

\begin{figure}
	\ContinuedFloat
	\begin{subfigure}{\columnwidth}
		\begin{tikzpicture}
		\node(a){\includegraphics[width=\columnwidth]{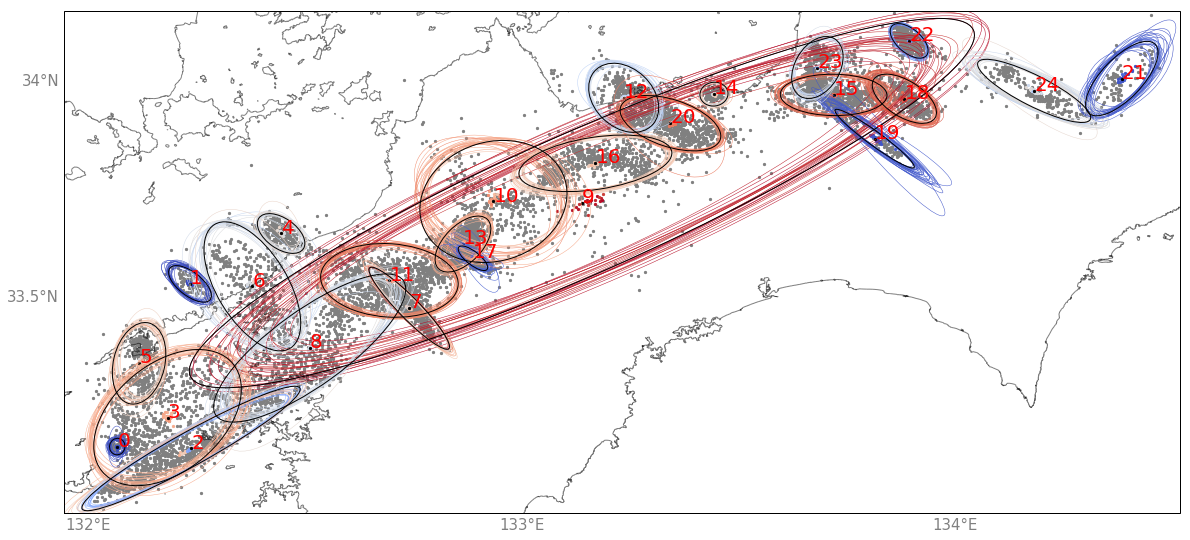}};
		\node at (a.south east)
		[
		anchor=center,
		xshift=-20mm,
		yshift=20mm
		]
		{
			\includegraphics[width=0.25\columnwidth]{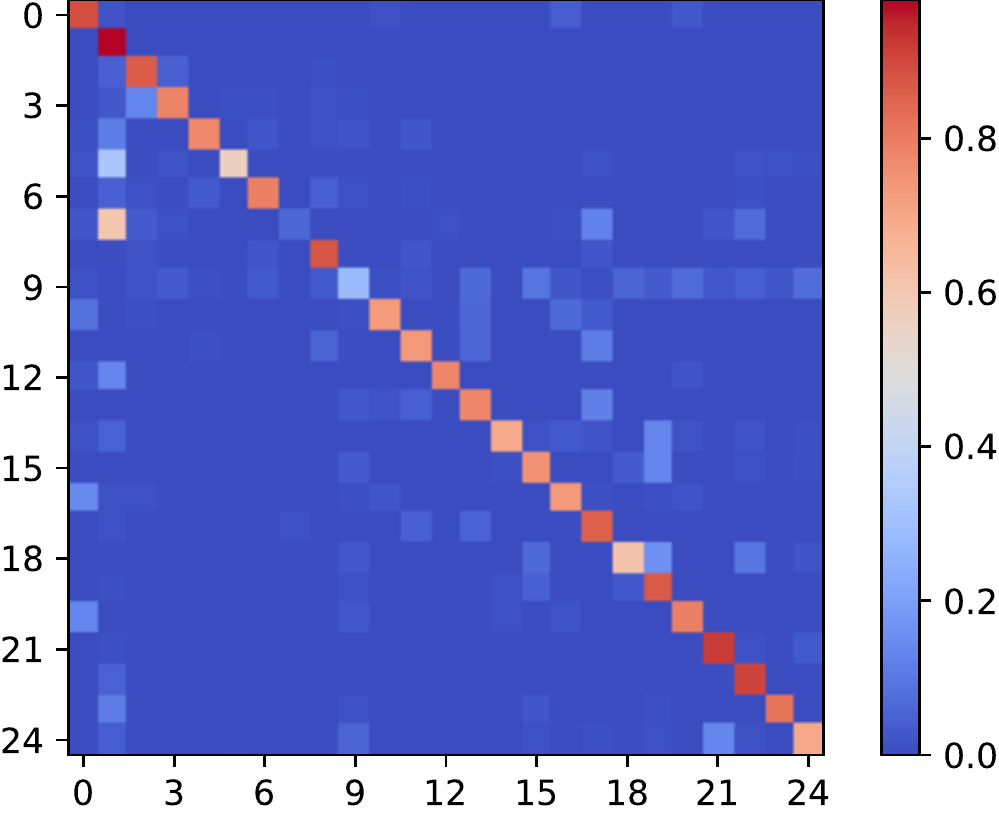}
		};
		
		\end{tikzpicture}%
		\caption{$K=25$}
	\end{subfigure}
	
	\begin{subfigure}{\columnwidth}
		\begin{tikzpicture}
		\node(a){\includegraphics[width=\columnwidth]{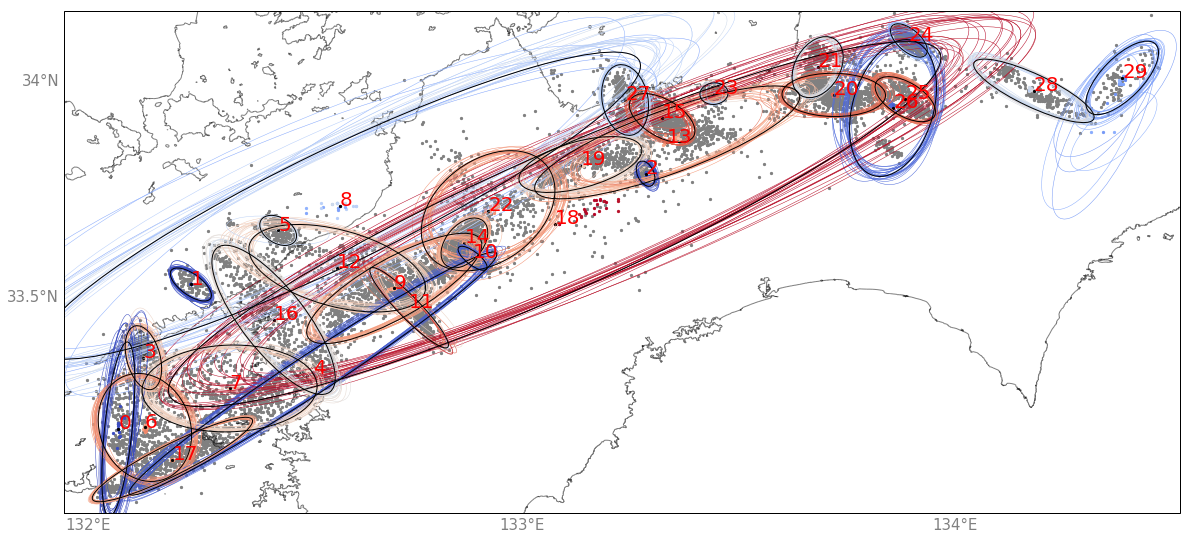}};
		\node at (a.south east)
		[
		anchor=center,
		xshift=-20mm,
		yshift=20mm
		]
		{
			\includegraphics[width=0.25\columnwidth]{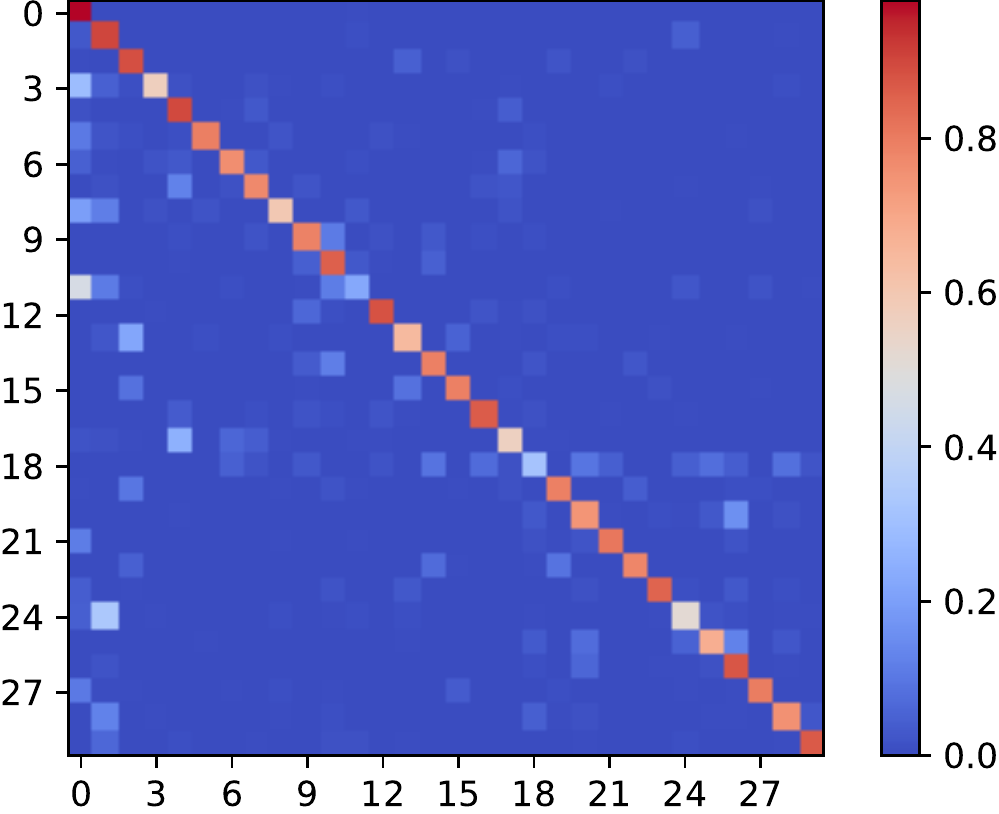}
		};
		
		\end{tikzpicture}%
		\caption{$K=30$}
	\end{subfigure}
	\caption{Posterior distributions of fitted models with number of hidden states $K=5,10,\dots,30$ for tremor occurrences in Shikoku region. Ellipses each map represent the 2D normal density of one hidden state for one sample from the posterior distribution. States are numbered in red. Colour of an ellipse indicate how likely a tremor will occur given the process is in the hidden state. In the bottom right corner of each map we give the mean transition matrix of the posterior distribution. Transition probabilities (array entries) and state probabilities (colour of ellipse) both use same colormap given in bottom right corner.  Furthermore grey dots represent the Shikoku tremor data points. Black ellipses and -dots represent mean parameters.}
	\label{fig:map}
\end{figure}

\subsection{Forecasting}
We carry out a Bayesian forecast from the model for a 5 day period (from December 11, 2012 to December 16, 2012). Note that the data for this period was excluded in the model fitting process. In order to forecast tremors we simulated 120 hourly datapoints (i.e for 5 days) from the model (with fixed number of hidden states $K=25$) for every 1000th MCMC sample (total of 500 simulations) of the approximate posterior distribution. Note that we used the same realization of the MCMC that was generated in the model fitting process (see previous section). We used the HMM simulator in the R package \textit{HMMextra0s} (freely available at \url{https://rdrr.io/cran/HMMextra0s/man/HMMextra0s-package.html}). 

\begin{figure}
	\begin{subfigure}{0.5\columnwidth}
		\includegraphics[width=\columnwidth]{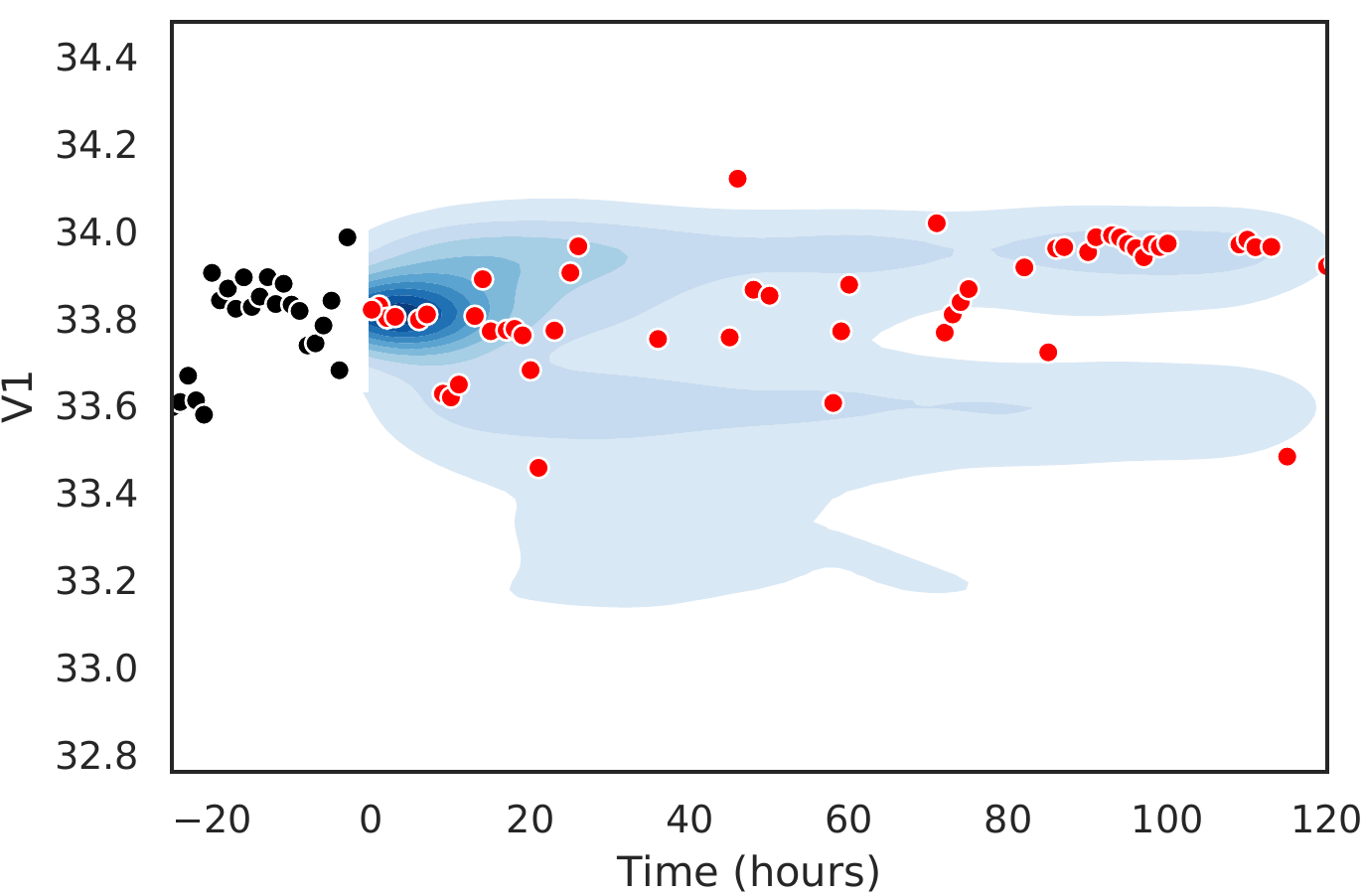}
		\caption{}
	\end{subfigure}
	\begin{subfigure}{0.5\columnwidth}
		\includegraphics[width=\columnwidth]{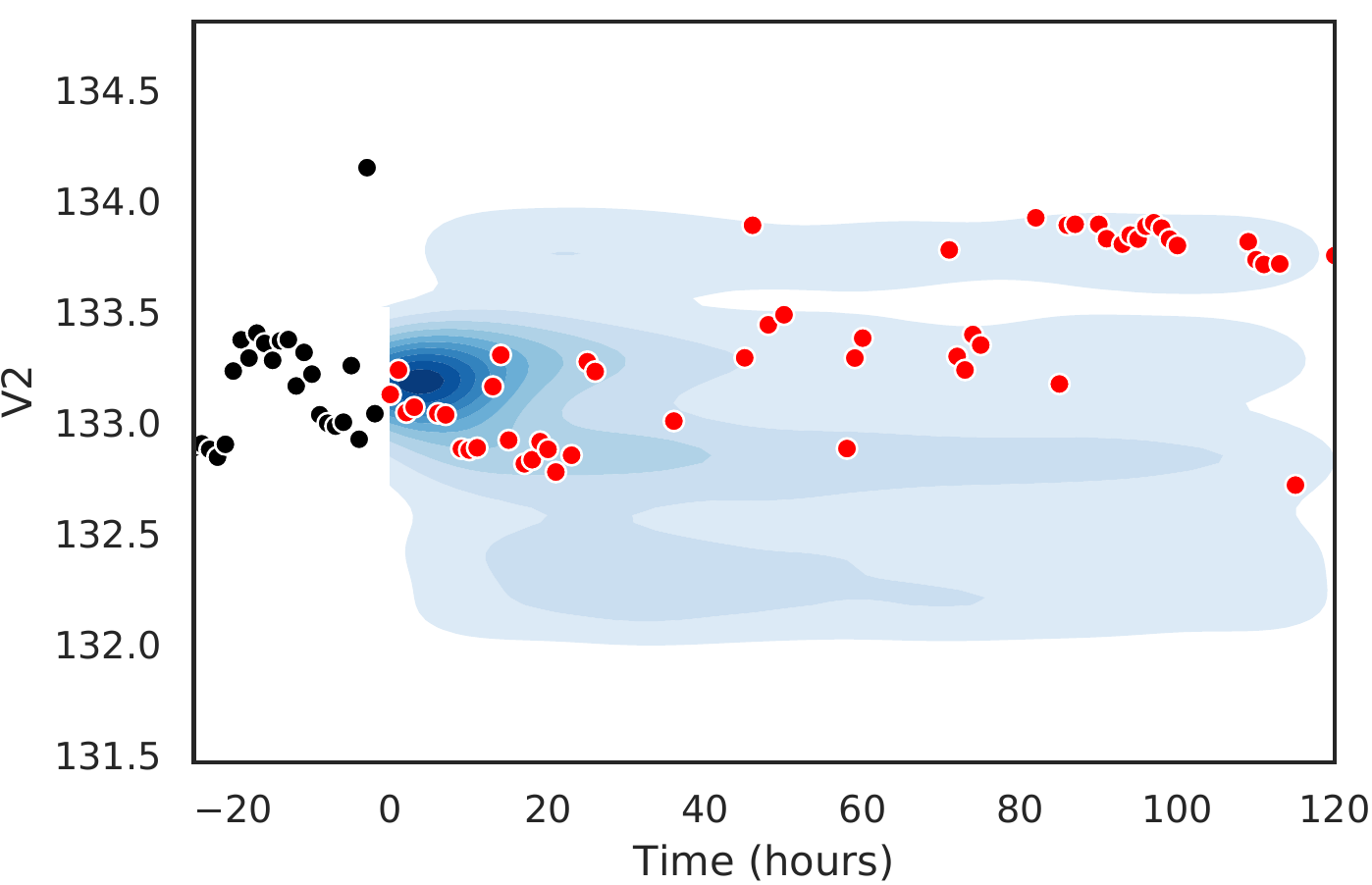}
		\caption{}	
\end{subfigure}
	\caption{We summarize the forecast simulations as two density plots. (a) Latitude predictions and data plotted against time (in hours). (b) Longitude predictions and data plotted against time (in hours). The red dots in both figures are the hourly Shikoku data for the time period from December 11, 2012 to December 30, 2012 (not included in data used for model fitting). Furthermore black dots in both figures are the hourly Shikoku data for the time period December 10, 2012 (included in data used for model fitting).}
	\label{fig:forecast}
\end{figure}

We summarize the 500 forecast simulations as a density in a longitude plot over time and a latitude plot over time (Figure \ref{fig:forecast}). Furthermore we plot the actual data as a scatterplot with red datapoints. We also include the last day  (December 10, 2012) of the data used for model fitting (as a scatterplot with black datapoints).   

We see that the model works well for the first two days. It captures nicely in which area the tremors occur. We also see that see that we get coverage from the forecast (density plot) for all the data points except for one outlier.  Furthermore we see that the variance in the model predictions increase with time. This is not unexpected since the further away our forecasts are from the present the less information our data contains about the future states of the process. It would be very unlikely to make an accurate forecast of more than a week.   

\begin{center}
	\section{Discussion}
\end{center}
In this paper we present an algorithm for evaluating HMM likelihoods that 
can run several orders of magnitude faster than the traditional Forward algorithm. Our algorithm requires more work, but the high level of parallization of the likelihood calculation translates into high data throughput. 

We have implemented the algorithm for an HMM model that categorises nonvolcanic tremor data. Furthermore we have integrated the algorithm as part of an R package for Bayesian analysis using the OpenCL framework with Python under the hood. It is however expected that a CUDA implementation for NVIDIA GPUs will achieve higher data throughput but this limits the algorithm to a single vendor. OpenCL on the other hand allows execution of the algorithm on any OpenCL compliant device such as Intel CPUs, AMD CPUs and GPUs, Qualcomm processors,  Xilinx FPGAs (Field-programmable gate array) and even NVIDIA GPUs. 

We have reported some runtime comparisons with implementations of the Forward algorithm. The efficieny gains in computation of the likelihood allowed us to conduct a detailed Bayesian analysis for tremor data of Shikoku region of Japan.  

Lastly, the OpenCL algorithm can be easily modified for other HMM models. In some cases only the evaluation function of the emission matrix needs to be updated. 

\begin{center}
	\item \section{Acknowledgement}
\end{center}	
Marnus Stolz received a doctoral scholarship from the NZ Marsden Fund (PIs David Bryant and Steven Higgins).

\bibliographystyle
{sysbio}
\bibliography{references}

\newpage
\appendix 
\begin{center}
	\item \section*{APPENDIX A: Formulas for prior distributions}
\end{center}

\subsection*{Symmetric Dirichlet distributions}
\[f(\gamma_1, \dots, \gamma_{K^2},\alpha) = \frac{\Gamma(\alpha K^2)}{\alpha^{K^2}} \prod_{i=1}^{K^2}\gamma_i^{\alpha-1} \qquad \text{, for } \alpha <1,\]
we note that probability mass is sparsely distributed among $\gamma_1, \dots, \gamma_{K^2}$ if $\alpha<1.$
\subsection*{Inverse-wishart distributions}
Suppose $\Psi$ is the scale matrix and $\nu$ the degrees of freedom then
\[f(\mathbf{x}, \Psi, \nu) = \frac{|\Psi|^{\nu/2}}{2^{\nu p/2}\Gamma_K(\frac{\nu}{2})} = |\mathbf{x}|^{-(\nu+K+1)/2} e^{-\frac{1}{2}tr(\Psi x^{-1})},\]
where $\Gamma_K$ is a multivariate gamma function 
\[\Gamma_K\left(\frac{\nu}{2}\right) =  \pi^{(\frac{\nu}{2})(\frac{\nu}{2}-1)/4} \prod^K_{j=1} \Gamma\left(\left(\frac{\nu}{2}\right)+(1-j)/2 \right).\]

\subsection*{Gamma distribution}
\[ f(x,\alpha,\beta) = \frac{\beta^\alpha x^{\alpha-1})e^{-\beta x}}{\Gamma(\alpha)}, \qquad \text{for $x>0$} \qquad \alpha, \beta >0.   \]

\begin{center}
	\item \section*{APPENDIX B: Additionals model fitted for nonvolcanic tremor data}
\end{center}
\begin{figure}[H]
	\centering
	\begin{subfigure}{\columnwidth}
		\begin{tikzpicture}
		\node(a){\includegraphics[width=\columnwidth]{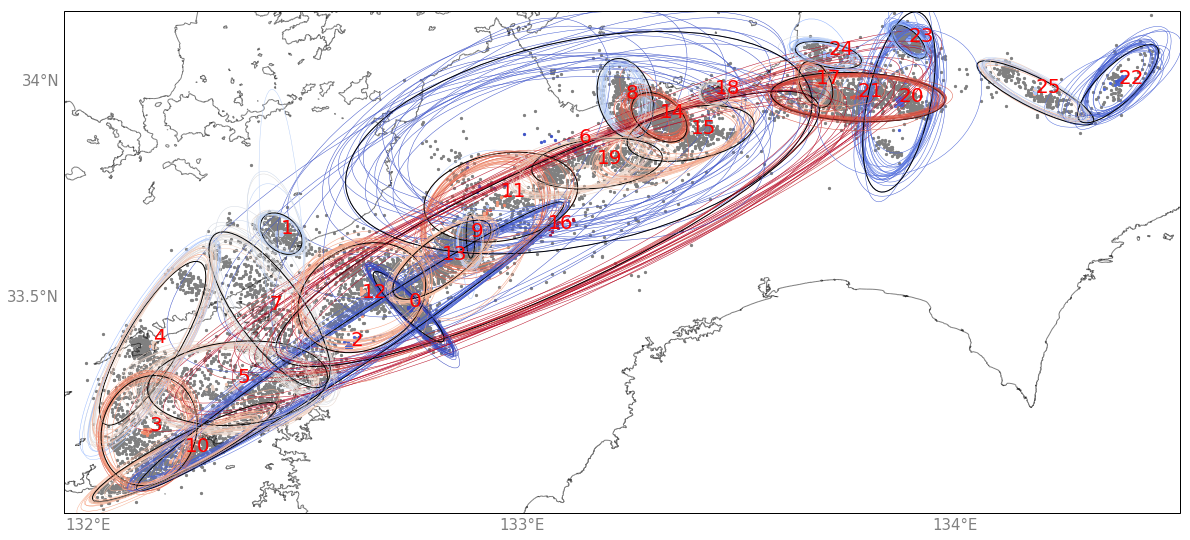}};
		\node at (a.south east)
		[
		anchor=center,
		xshift=-20mm,
		yshift=20mm
		]
		{
			\includegraphics[width=0.25\columnwidth]{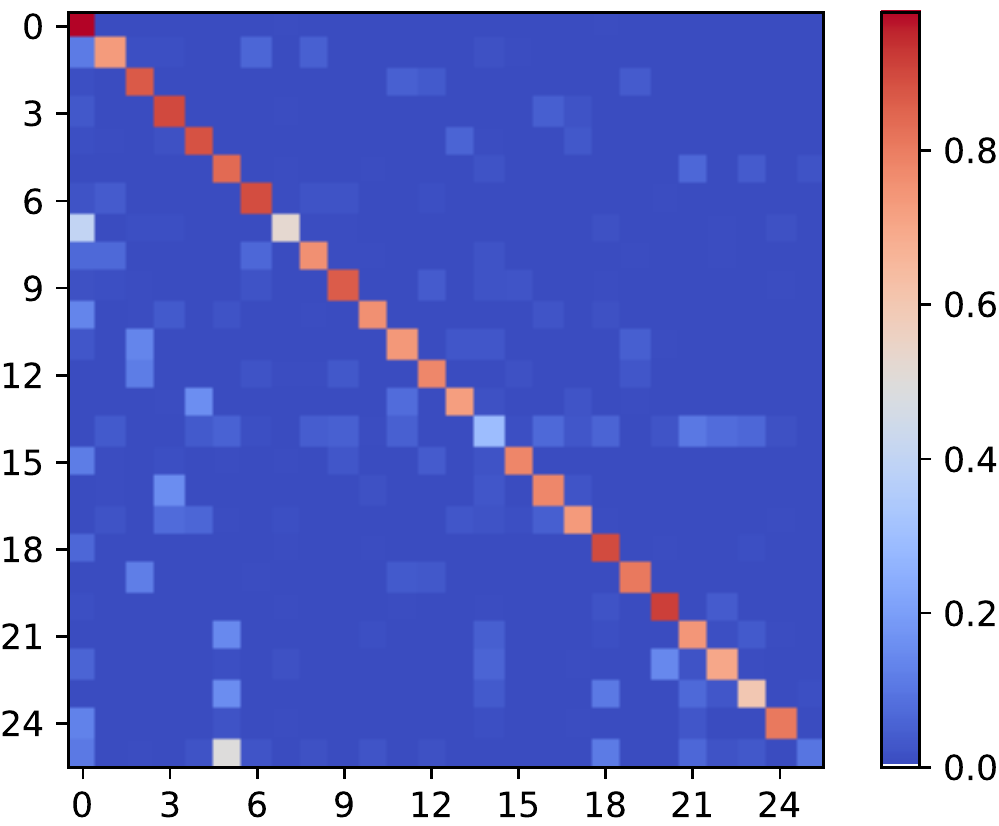}
		};
		
		\end{tikzpicture}%
		\caption{$K=26$}
	\end{subfigure}
	
	\begin{subfigure}{\columnwidth}
		\begin{tikzpicture}
		\node(a){\includegraphics[width=\columnwidth]{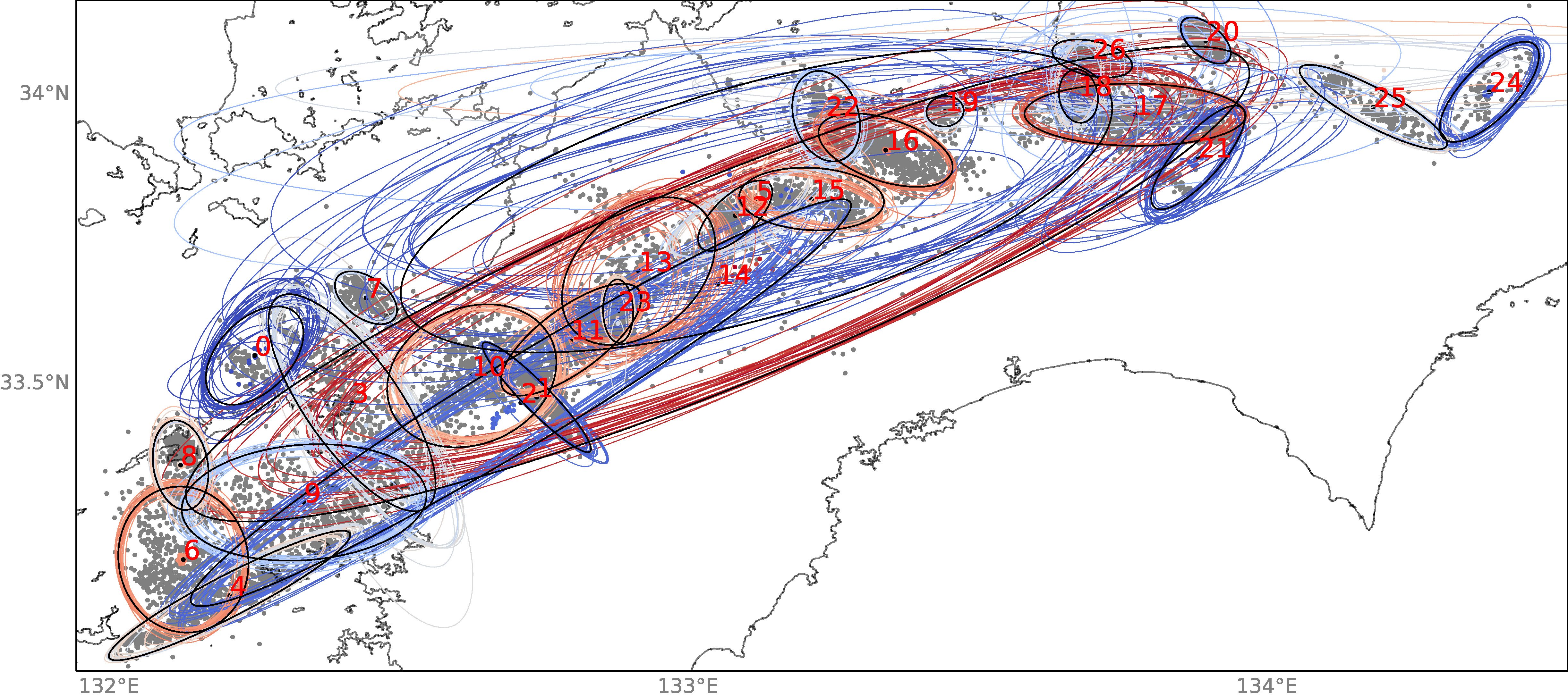}};
		\node at (a.south east)
		[
		anchor=center,
		xshift=-20mm,
		yshift=20mm
		]
		{
			\includegraphics[width=0.25\columnwidth]{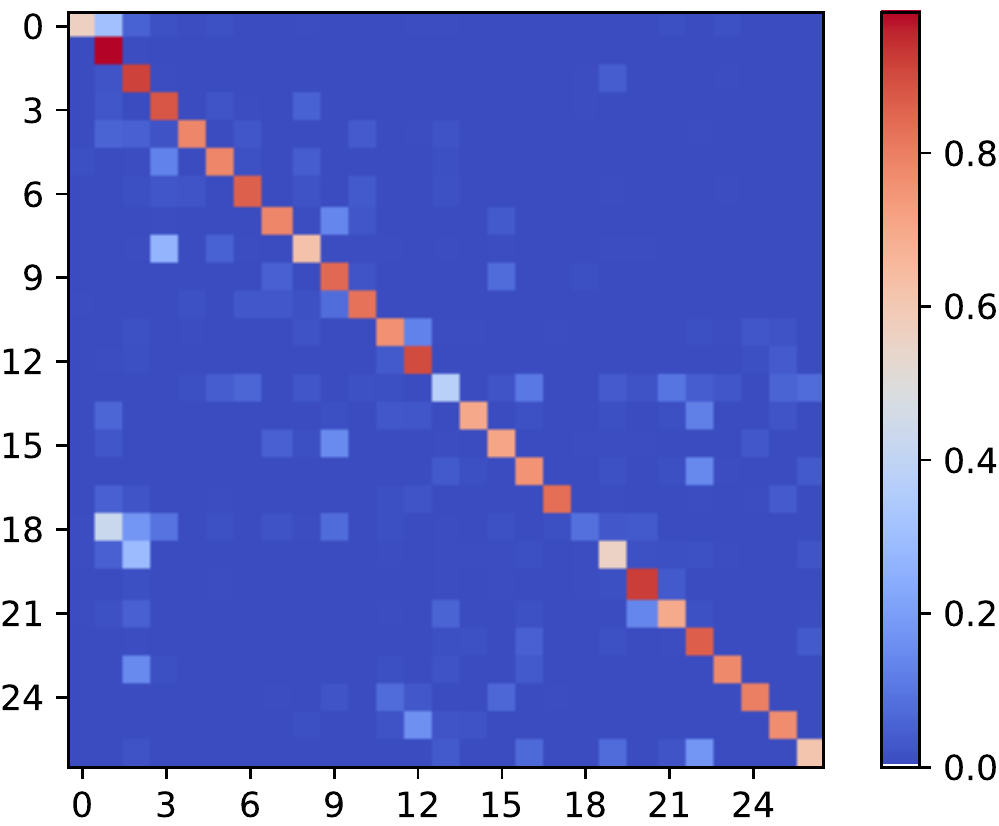}
		};
		
		\end{tikzpicture}%
		\caption{$K=27$}
	\end{subfigure}
	\caption{Posterior distributions of fitted models with number of hidden states $K=26,27$ for tremor occurrences in Shikoku region. Ellipses each map represent the 2D normal density of one hidden state for one sample from the posterior distribution. States are numbered in red. Colour of an ellipse indicate how likely a tremor will occur given the process is in the hidden state. In the bottom right corner of each map we give the mean transition matrix of the posterior distribution. Transition probabilities (array entries) and state probabilities (colour of ellipse) both use same colormap given in bottom right corner.  Furthermore grey dots represent the Shikoku tremor data points. Black ellipses and -dots represent mean parameters.}
	\label{fig:map-append}
\end{figure}
\begin{center}
	\item \section*{APPENDIX C: Tabulated posterior statistics for number of hidden states K=25}
\end{center}
\begin{center}
\begin{longtable}{lllllll}
		\caption{Combined GPU-hmmer parameter summary after 5,000,000 MCMC iterations for soybean dataset} \\
		\hline
		\textbf{hmmer} & \textbf{mean} & \textbf{variance} & \textbf{HPD} & \textbf{ACT} & \textbf{ESS}\\ \hline
		\endfirsthead
		\multicolumn{4}{c}%
		{\tablename\ \thetable\ -- \textit{Continued from previous page}} \\
		\hline
		\textbf{hmmer} & \textbf{mean} & \textbf{variance} & \textbf{HPD} & \textbf{HPD} & \textbf{ACT} & \textbf{ESS}\\
		\hline
		\endhead
		\hline \multicolumn{4}{r}{\textit{Continued on next page}} \\
		\endfoot
		\hline
		\endlastfoot 
		\hline
		posterior & -1.0987E+03 & 7.6624E+04 & -1.6110E+03 & -4.0624E+02 & 35830.8997  & 239.5444  \\
		Gamma1    & 8.6710E-01  & 2.4186E-04 & 8.3770E-01  & 8.9248E-01  & 29571.9564  & 169.0791  \\
		Gamma2    & 1.6289E-02  & 4.7111E-05 & 3.8358E-03  & 2.8083E-02  & 14455.9226  & 345.8790  \\
		Gamma3    & 1.8235E-03  & 8.1549E-06 & 5.3391E-06  & 5.9633E-03  & 31080.9139  & 160.8704  \\
		Gamma4    & 1.8604E-03  & 9.1692E-06 & 8.0933E-06  & 6.5181E-03  & 23964.1982  & 208.6446  \\
		Gamma5    & 1.4205E-03  & 1.2756E-06 & 4.4740E-05  & 3.3164E-03  & 14213.5729  & 351.7764  \\
		Gamma6    & 3.4741E-03  & 7.6753E-06 & 2.8037E-05  & 8.2833E-03  & 21277.4832  & 234.9902  \\
		Gamma7    & 2.9676E-03  & 3.5263E-06 & 4.7983E-05  & 5.7603E-03  & 15272.0127  & 327.3963  \\
\end{longtable}
\end{center}

\end{document}